\documentclass[acus]{JAC2003}

\usepackage{fancyhdr, graphicx}
\usepackage{float}
\usepackage{color}
\usepackage{amsmath,amssymb,wasysym}

\title{LHC Optics Determination with Proton Tracks Measured in the Roman Pot Detectors of the TOTEM Experiment}
\author{
	F. Nemes, E\"otv\"os University, Budapest, Hungary\\
	H. Niewiadomski, H. Burkhardt, CERN, Geneva, Switzerland \\ on behalf of the TOTEM collaboration
}

\let\olditemize=\itemize
\def\itemize{
\olditemize
  \setlength{\itemsep}{5pt}
  \setlength{\parskip}{-5pt}
}

\begin{document}
\date{}
\maketitle

\begin{abstract}
The TOTEM experiment at the LHC is equipped with near beam movable devices -- called Roman Pots (RP) -- which detect protons scattered at the interaction point (IP5) 
arriving to the detectors through the magnet lattice of the LHC. Proton kinematics at IP5 is reconstructed from positions and angles measured by the RP detectors, 
on the basis of the transport matrix between IP5 and the RP locations. The precision of optics determination is therefore of the key importance for the experiment. 
TOTEM developed a novel method of machine optics determination making use of angle-position distributions of elastically scattered protons observed in the RP detectors
together with the data retrieved from several machine databases. 
The method has been successfully applied to the data samples registered in 2010 and 2011. The studies show that the transport matrix could be estimated with a 
precision better than 1\%.
\end{abstract}

\section{The Roman Pots of the TOTEM experiment}
Proton-proton elastic scattering was measured by the TOTEM experiment at the CERN Large Hadron Collider at $\sqrt{s}$ = 7 TeV in dedicated runs~\cite{Antchev:2011zz,Antchev:2011vs}. 
To detect leading protons scattered at angles as small as 1$\mu$rad, silicon sensors are placed in movable beam-pipe insertions, —so-called “Roman Pots” (RP), located 
symmetrically on either side of the LHC intersection point IP5 at distances up to $220\,$m from it.  
Each RP station is composed of two units separated by a distance of about $5\,$m. A unit consists of 3 RPs, two approaching the outgoing beam vertically and 
one horizontally, allowing for a partial overlap between horizontal and vertical detectors and the alignment precision of 10 $\mu$m.  

\section{Proton transport from IP5 to the Roman Pots}
Scattered protons are detected in the Roman Pots after having moved through a segment of the LHC lattice containing 29 magnets per beam.
The trajectory of protons with transverse positions\footnote{The $*$ superscript indicates the LHC Interaction Point 5} $(x^*,y^*)$ and angles $(\Theta_x^*, \Theta_y^*)$ at IP5 are described with a linear formula  
\begin{align}
	\vec{d}=T\cdot\vec{d}^{*},  
	\label{proton_trajectories}
\end{align} 
where $\vec{d}=\left\{x,\Theta_x,y,\Theta_y,\Delta p/p\right\}^{T}$ with nominal beam momentum $p$  
and momentum loss $\Delta p$. The transport matrix $T$ is defined by the optical functions
\begin{align}
	T=\left(
		\begin{array}{ccccc} 
			v_x         & L_x     & m_{13}   & m_{14}  & D_x  \\
			v_x'        & L_x'    & m_{23}   & m_{24}  & D_x' \\
			m_{31}      & m_{32}  & v_y      & L_y     & D_y  \\
			m_{41}      & m_{42}  & v_y'     & L_y'    & D_y' \\ 
			0           & 0       & 0        & 0       & 1
    	\end{array}  
	\right).
	\label{transport_matrix}
\end{align}
The \textit{magnification} $v_{x,y}=\sqrt{\beta_{x,y}/\beta^*}\cos\Delta\phi_{x,y}$ and the  {\it effective length} $L_{x,y}=\sqrt{\beta_{x,y}\beta^*}\sin\Delta\phi_{x,y}$ are functions 
of the {\it betatron amplitude} $\beta_{x,y}$ and the relative {\it phase advance} $\Delta\phi_{x,y}=\int^{\text{\tiny RP}}_{\text{\tiny IP}}\beta(s)_{x,y}^{-1}ds$ and are 
particularly important for the proton kinematics reconstruction. The coupling coefficients $m_{i,j}$ are close to 0 and the vertex contributions are canceled due to the 
anti-symmetry of the scattering angles.  Therefore, the kinematics of elastically 
scattered protons at IP5 can be reconstructed from Equation~(\ref{proton_trajectories}) as:
\begin{align}
	\Theta_{y}^{*} \approx \frac{y_{\text{\tiny RP}}}{L_{y,\text{\tiny RP}}}\;\;\;\;\;\;
	\Theta_{x}^{*} \approx \frac{1}{\frac{dL_{x,\text{\tiny RP}}}{ds}}\left(\Theta_{x,\text{\tiny RP}}-\frac{dv_{x,\text{\tiny RP}}}{ds}x^{*}\right),
\end{align}
where ``RP" defines the measurement location. As the values of the reconstructed angles are directly inversely proportional to the optical functions, the 
accuracy of optics defines the systematic errors of the final physics results.\par
The proton transport matrix $T\left(s;\mathcal{M}\right)$ over a distance of $s$  is defined by the machine settings $\mathcal{M}$. 
It is calculated with the MAD-X \cite{MADX} code for each group of runs with identical optics based on several data sources. The magnet currents are retrieved from
TIMBER \cite{TIMBER} and are converted 
to strengths with LSA \cite{LSA}, which implements the conversion curves measured by FIDEL \cite{FIDEL}. The WISE database \cite{WISE} contains the 
measured imperfections (field harmonics, magnet displacements and rotations).\par
However, the lattice is subject to additional $\Delta \mathcal{M}$ \textit{imperfections}, not measured well enough so far, which alter the
transport matrix by $\Delta T$:
\begin{align*}
	T\left.(s;\, \mathcal{M}\right) \rightarrow T\left.(s;\, \mathcal{M}+\Delta \mathcal{M}\right) = T\left.(s;\, \mathcal{M}\right)+\Delta T .
\end{align*}
The 5--10\% precision of $\Delta \beta/\beta$ beating measurement does not allow to estimate $\Delta T$ with the accuracy required by the TOTEM physics program.
However, the magnitude of $\left|\Delta T\right|$ can be evaluated from the tolerances of the LHC imperfections of which the most important are:
\begin{itemize}
	\item Strength conversion error $I \rightarrow B\,, \sigma(B)/B\approx10^{-3}$
	\item Beam momentum offset $\sigma(p)/p \approx 10^{-3}\,$.
\end{itemize}
Their impact on optical functions is presented in Table $\ref{Lysensitivity}$. 
It is clearly visible that the imperfections of the inner triplet (the MQXA and MQXB magnets) are of high influence on the transport matrix while 
the optics is less sensitive to the quadrupoles MQY and MQML.
 
Other imperfections are of lower significance:
\begin{itemize}
	\item Magnet rotations $\sigma(\phi)\approx1$ mrad
	\item Beam harmonics $\sigma(B)/B \approx 10^{-4}$
	\item Power converter errors $\sigma(I)/I \approx 10^{-4}$
	\item Magnet positions $\Delta x,\Delta y \approx 100\,\mu$m.
\end{itemize}
Generally, as can be seen in Table~\ref{Lysensitivity}, for large $\beta^*$ optics the magnitude of $\Delta T$ is sufficiently small from the viewpoint of 
data analysis and therefore $\Delta T$ does not need to be precisely estimated. However, the low $\beta^*$ optics sensitivity to the machine imperfections
is significant and cannot be neglected. Fortunately, in this case $\Delta T$ can be determined precisely enough from the proton tracks in the Roman Pots.
\begin{table}[hbt]
    \begin{center}
        \begin{tabular}{  c | c | c |} \cline{2-3}
               											& \multicolumn{2}{|c|}{\bf $\mathbf{\boldsymbol\delta L_y/L_y}$\,[\%]}      \\ \hline
            \multicolumn{1}{|c|}{\bf Perturbed element}   	& $\mathbf{\boldsymbol\beta^{*}=3.5\,m}$	& $\mathbf{\boldsymbol\beta^{*}=90\,m}$	\\ \hline
            \multicolumn{1}{|c|}{MQXA.1R5}        		& $\phantom-0.98$            				& $\phantom-0.14$    						\\
            \multicolumn{1}{|c|}{MQXB.A2R5}       		& $-2.24$            					& $-0.23$    						\\
            \multicolumn{1}{|c|}{MQXB.B2R5}       		& $-2.42$            					& $-0.25$   						\\
            \multicolumn{1}{|c|}{MQXA.3R5}        		& $\phantom-1.45$       	     			& $\phantom-0.20$    						\\
            \multicolumn{1}{|c|}{MQY.4R5.B1}      		& $-0.10$           	  				& $-0.01$    						\\
            \multicolumn{1}{|c|}{MQML.5R5.B1}     		& $\phantom-0.05$             				& $\phantom-0.04$    						\\
            \multicolumn{1}{|c|}{$\Delta$p/p}     		& $-2.19$             					& $\phantom-0.01$   						\\ \hline
        \end{tabular}
        \caption{Sensitivity of the vertical effective length $L_y$ to magnet strengths and beam momentum perturbed by 1~$\permil$ for low- and large-$\beta^*$ optics. 
}
		\label{Lysensitivity}
    \end{center}
\end{table}
\vspace{-20pt}
\section{ Constraints from proton tracks in the Roman Pots}

The elements of the transport matrix are functions of the betatron amplitudes $\beta_{x,y}$ and the phase advances $\phi_{x,y}$. Therefore they are mutually 
related. Moreover, the elastic scattering ensures that the scattering angles in both arms are identical: 
\begin{align}
    \Theta^{*}_{x,b_1} = \Theta^{*}_{x,b_2}\,,\; 
    \Theta^{*}_{y,b_1} = \Theta^{*}_{y,b_2}\,, 
    \label{collinearity_cut}
\end{align}
which allows to compute ratios between the effective lengths of the two beams. From Equation (\ref{proton_trajectories}) we get: 
\begin{align}
    R_{1}&\equiv\frac{\Theta_{x,b_1,\text{RP}}}{\Theta_{x,b_2,\text{RP}}} \approx \frac{\frac{dL_{x,b_1,\text{RP}}}{ds}\Theta^{*}_{x,b_1}}{\frac{dL_{x,b_2,\text{RP}}}{ds}\Theta^{*}_{x,b_2}} = 
        \frac{\frac{dL_{x,b_1,\text{RP}}}{ds}}{\frac{dL_{x,b_2,\text{RP}}}{ds}}\,, \\
    \label{ratiodLxdsandratioLy}
    R_{2}&\equiv\frac{y_{b_1,\text{RP}}}{y_{b_2,\text{RP}}} \approx \frac{L_{y,b_1,\text{RP}}}{L_{y,b_2,\text{RP}}},
\end{align}
where $b_1$ and $b_2$ indicate beam 1 and beam 2. The ratios $R_1$ and $R_2$ can be estimated with a 0.5\% precision.

Furthermore, the distributions of proton angles and positions detected in Roman Pots define ratios of 
certain elements of the transport matrix $T$. First of all, $dL_y/ds$ and $L_y$ are related by 
\begin{align}
	R_3\equiv\frac{\Theta_{y,b_1,\text{RP}}}{y_{b_1,\text{RP}}} \approx \frac{\frac{dL_{y,b_1,\text{RP}}}{ds}}{L_{y,b_1,\text{RP}}}\,,\;
\end{align} 
with a 0.5\% precision, and $R_4$ is the same for beam 2.

Similarly, we exploit the horizontal distributions to quantify the relation between $dL_x/ds$ and $L_x$. Contrary to the previous case, $L_x$ is
close to $0$ and instead of defining the ratio we rather estimate the position $s$ (with the precision of about $1\,$m) along the beam where $L_x$ equals to $0$ by solving 
\begin{align}
	\frac{L_x(s)}{dL_x(s_1)/ds}= \frac{L_x(s_1)}{dL_x(s_1)/ds} + \left(s-s_1\right)=0\,,
\end{align}
where $s_1$ is the beginning of the Roman Pot station. The ratio $\frac{dL_{x}(s_1)}{ds}/L_{x}(s_1)$ is defined by the proton distributions  
\begin{align}
    R_5\equiv\frac{x_{b_1,\text{RP}}}{\Theta_{x,b_1,\text{RP}}} \approx \frac{L_{x,b_1,\text{RP}}}{\frac{dL_{x,b_1,\text{RP}}}{ds}}\,,\;
    R_6\equiv\frac{x_{b_2,\text{RP}}}{\Theta_{x,b_2,\text{RP}}} \approx \frac{L_{x,b_2,\text{RP}}}{\frac{dL_{x,b_2,\text{RP}}}{ds}}.
\end{align}
Finally, tracks determine as well the coupling components of $T$. Due to $L_x\approx0$ at the Roman Pot locations, the further four constraints can
be defined
\begin{align}
	R_7\equiv\frac{x_{b_1,\text{near pots}}}{y_{b_1,\text{near pots}}}\approx\frac{m_{14,b_1,\text{near pots}}}{L_{y,b_1,\text{near pots}}}\,,	
\end{align}
$R_8$ is defined with the far pots, and $R_{9,10}$ respectively for beam 2. These four constraints can be estimated with a 3\% accuracy.

\section{Optics matching}
On the basis of the constraints $R_1...R_{10}$, $\Delta T$ can be determined with the $\chi^2$ minimization procedure. The relevant lattice imperfections were selected forming a 26 dimensional 
optimization phase space, which includes the magnet strengths, rotations and beam momenta. Due to the high dimensionality of the phase space and approximately linear structure of the 
problem there is no unique solution. Therefore, the optimization is subject to additional constraints defined by the machine tolerances. Finally, the  
$\chi^2$ is composed of the part defined by the values measured with the Roman Pots (discussed in the previous section) and such reflecting the LHC tolerances:
\begin{align}
\chi^2 = \chi_\text{Measured}^2 + \chi_\text{Design}^2,
\end{align}
where the design part  
\begin{align*}
                                        \chi_\text{Design}^2 = &\sum_{i=1}^{12}\left(\frac{k_{i}-k_{i,\text{{\tiny MADX}}}}{\sigma(k_i)}\right)^2 + 
                                        \sum_{i=1}^{12}\left(\frac{\phi_{i}-\phi_{i,\text{\tiny MADX}}}{\sigma(\phi_i)}\right)^2 +\\
                                        &\sum_{i=1}^{2}\left(\frac{p_{i}-p_{i,\text{\tiny MADX}}}{\sigma(p_i)}\right)^2\,,
\end{align*}
defines the nominal machine as an attractor in the phase space, and the measured part
\begin{align}
    \chi^2_\text{Measured} = \sum_{i=1}^{10}\left(\frac{R_{i}-R_{i,\text{\tiny MADX}}}{\sigma(R_i)}\right)^2
\end{align} 
contains the track based constraints $R_1...R_{10}$ together with their errors. The subscript ``MADX" defines the parameter optimized with the MAD-X software. 

Table \ref{matching_result} presents the results of the optimization procedure for $\beta^*=3.5\,$m. The obtained value of the effective length $L_y$ of beam 1 
is close to the nominal one, while beam 2 shows a significant change. The same pattern applies to the values of $dL_x/ds$. 
\begin{table}[hbt]\renewcommand{\arraystretch}{1.0}\addtolength{\tabcolsep}{-4pt}
\begin{center}
	\begin{tabular}{ c |c|c|c|c|}\cline{2-5}
											&	$\mathbf{L_{y,b_1}}$[m]  	&	$\mathbf{dL_{x,b_1}/ds}$			& {\bf $\mathbf{L_{y,b_2}}$[m]} 	&   $\mathbf{dL_{x,b_2}/ds}$			\\\hline		
	\multicolumn{1}{|c|}{\bf Nominal}		& 	$22.4$ 						&	$-3.21\cdot$10$^{-1}$ 				& $18.4$ 						&	$-3.29\cdot$10$^{-1}$ 	\\\hline
	\multicolumn{1}{|c|}{\bf Matched}		& 	$22.6$						& 	$-3.12\cdot$10$^{-1}$ 				& $20.7$						& 	$-3.15\cdot$10$^{-1}$	\\\hline
	\end{tabular}
\end{center}
	\caption{Selected optical functions of both LHC beams obtained with the matching procedure compared to their nominal values for $\beta^*=3.5\,$m.}
	\label{matching_result}
\end{table}
\vspace{-15pt}
\section{Monte-Carlo validation}
The procedure has been extensively verified with Monte Carlo studies. The nominal machine settings were perturbed in order
to simulate the LHC imperfections and the simulated proton tracks were used afterwards to calculate the optimization 
constraints $R_1...R_{10}$. The study included the impact of 
	\begin{itemize}
		\item magnet strengths 
		\item beam momenta
		\item displacements, rotations
		\item kickers, harmonics
		\item elastic scattering $\Theta$-distributions
	\end{itemize}\par
\begin{figure}[htb]
	\includegraphics[trim = 0mm 12mm 0mm 12mm, clip, width=0.45\textwidth]{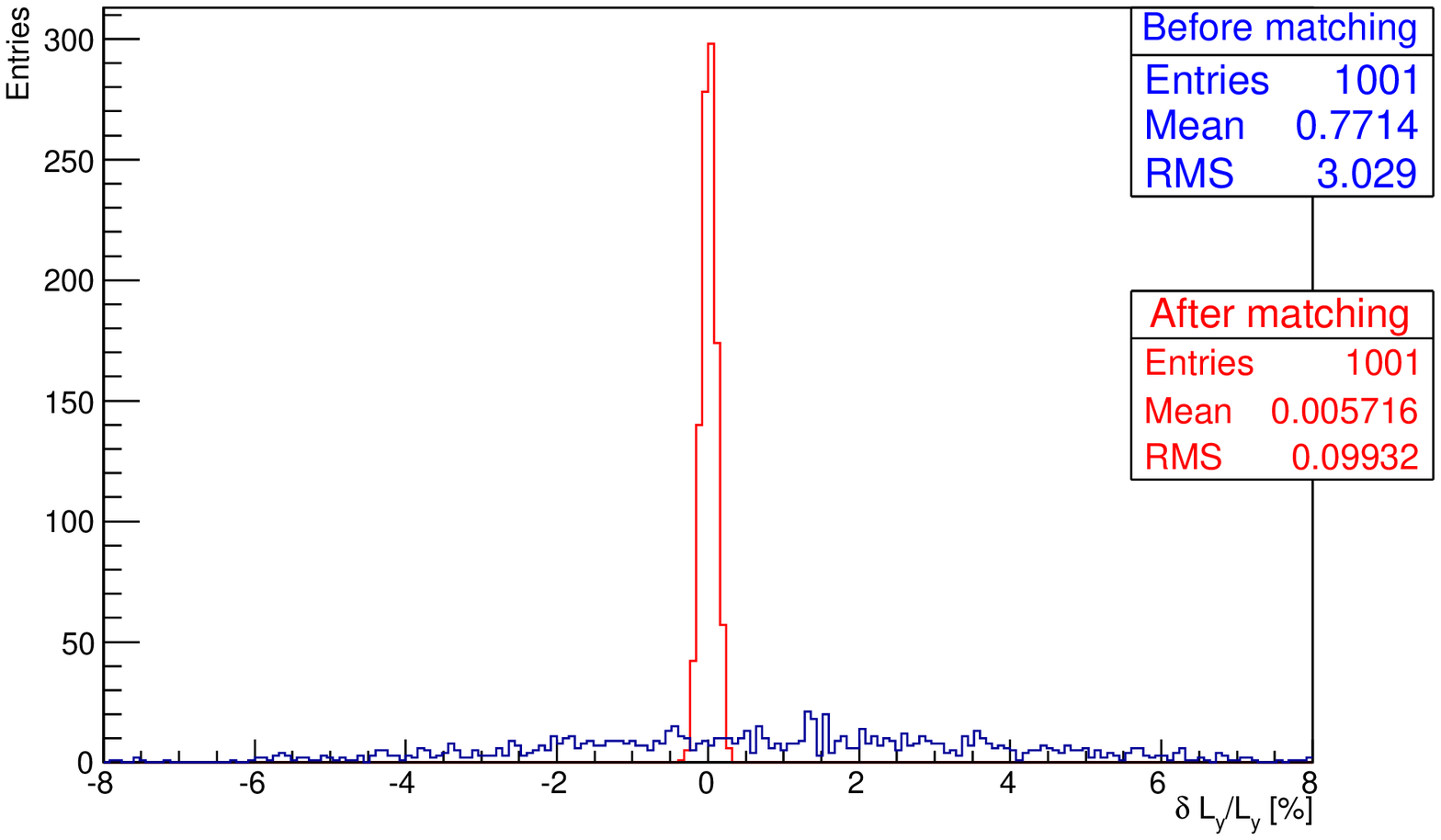}
	\caption{Relative error distribution of $L_{y}$ for beam 1 before and after matching.}
	\label{MCresultLy}
\end{figure}
\begin{figure}[htb]
    \includegraphics[trim = 0mm 12mm 0mm 12mm, clip, width=0.45\textwidth]{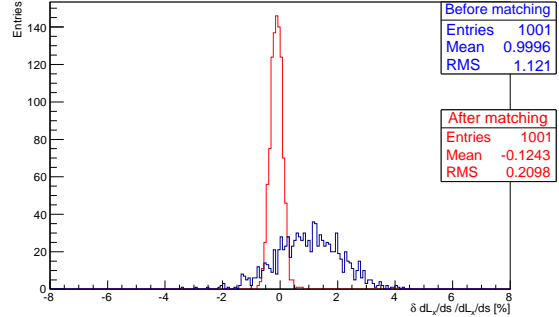}
	\caption{Relative error distribution of $dL_{x}/ds$ for beam 1 before and after matching.}
	\label{MCresultdlxds}
\end{figure}

The results obtained for the $\beta^*=3.5\,$m study are summarized in Figures \ref{MCresultLy} and \ref{MCresultdlxds}
and their statistical description is given in Table \ref{MCestimations}. The distributions of optical functions' errors indicate that the optical functions can be reconstructed with
a precision of 0.2\%, which confirms the validity of the proposed approach.
\begin{center}
\begin{table}[hbt]\renewcommand{\arraystretch}{1.0}\addtolength{\tabcolsep}{-4pt}
   \begin{tabular}{  c | c | c | c | c | c | c |} \cline{2-5}
       & \multicolumn{4}{|c|}{\bf Machine with imperfections} \\ 
       & \multicolumn{2}{|c|}{\bf before} & \multicolumn{2}{|c|}{\bf after matching} \\ \hline
       \multicolumn{1}{|c|}{\bf Optical function }                     					& {\bf Mean}   & {\bf RMS}     	& {\bf Mean}    				& {\bf RMS}   \\ 
       \multicolumn{1}{|c|}{\bf relative error}                        					&    [\%]      &     [\%]      	&     [\%]        				& [\%]  \\ \hline
       \multicolumn{1}{|c|}{ $\frac{\delta L_{y,b_1}}{L_{y,b_1}}\;$ }              	&  $0.77$         & $3.0$           & $\phantom-5.7\cdot10^{-3}$    & $9.9\cdot10^{-2}$ \\   
       \multicolumn{1}{|c|}{ $\frac{\delta dL_{x,b_1}/ds}{dL_{x,b_1}/ds}\;$ }       &  $1.0$          & $1.1$           & $-1.2\cdot10^{-1}$         	& $2.1\cdot10^{-1}$  \\\hline    
        \multicolumn{1}{|c|}{ $\frac{\delta L_{y,b_2} }{L_{y,b_2}} \;$ }            &  $2.0$          & $3.8$           & $\phantom-1.5\cdot10^{-1}$    & $9.5\cdot10^{-2}$ \\
        \multicolumn{1}{|c|}{ $\frac{\delta dL_{x,b_2}/ds}{dL_{x,b_2}/ds}\;$ }      & $-1.14$         & $1.2$           & $-7.6\cdot10^{-2}$       		& $2.1\cdot10^{-1}$  \\\hline

   \end{tabular}
	\caption{Monte-Carlo validation results of Roman Pot track based optics estimation. The machine imperfections induce large spread of optical functions. 
            The matching procedure estimates the optics with errors lower than 2.1 $\permil$.} 
	\label{MCestimations}
	\end{table}
\end{center}
\section{Conclusions and outlook}
TOTEM proposed a novel approach to optics estimation. First of all, the method allows to asses the optical functions' errors from 
machine tolerances. Secondly, it allows to determine the real optics solely from the Roman Pot proton tracks. The method has been validated with
the Monte Carlo studies both for large- and low-$\beta^*$ optics. With its application TOTEM has published elastic scattering distributions
obtained with different running conditions. It is foreseen to extend the proposed approach to model the transport of protons with large momentum
loss.

\vfill

\end{document}